\documentclass[epj]{svjour}
\usepackage{graphics}
\usepackage{bm}
\usepackage{amsfonts}
\usepackage{latexsym}
\usepackage[latin1]{inputenc}
\usepackage{graphicx}
\usepackage{amsmath}
\usepackage{palatino}
\usepackage{mathpazo}
\usepackage{textcomp}
\usepackage{float}
\usepackage{booktabs}
\usepackage{dcolumn}
\usepackage{ragged2e}
\usepackage{hyperref}
\hypersetup{colorlinks,citecolor=blue}
\usepackage{amsmath}
\usepackage{xcolor}
\usepackage{orcidlink}
\usepackage[caption=false]{subfig}
\usepackage{commath}
\begin{document}
\title{Power-law cosmology in Weyl-type $f(Q,T)$ gravity}
\author{Gaurav Gadbail\orcidlink{0000-0003-0684-9702}, Simran Arora\orcidlink{0000-0003-0326-8945} \and P.K. Sahoo\orcidlink{0000-0003-2130-8832}}
%
%
\institute{Department of Mathematics, Birla Institute of Technology and
Science-Pilani, Hyderabad Campus, Hyderabad-500078, India.\\
E-mail: G. Gadbail: gauravgadbail6@gmail.com, S. Arora: dawrasimran27@gmail.com, P.K. Sahoo: pksahoo@hyderabad.bits-pilani.ac.in }

\date{Received: 27th  July 2021 / Revised version: 30th Sept. 2021}
%
\abstract{
Gravity is attributed to the spacetime curvature in classical General Relativity (GR). But, other equivalent formulation or representations of GR, such as torsion or non-metricity have altered the perception. We consider the Weyl-type $f(Q, T)$ gravity, where $Q$ represents the non-metricity and $T$ is the trace of energy momentum temsor, in which the vector field $\omega_{\mu}$ determines the non-metricity $Q_{\mu \nu \alpha}$ of the spacetime. In this work, we employ the well-motivated $f(Q, T)= \alpha Q+ \frac{\beta}{6k^{2}} T$, where $\alpha$ and $\beta$ are the model parameters. Furthermore, we assume that the universe is dominated by the pressure-free matter, i.e. the case of dust ($p=0$). We obtain the solution of field equations similar to a power-law in Hubble parameter $H(z)$. We investigate the cosmological implications of the model by constraining the model parameter $\alpha$ and $\beta$ using the recent 57 points Hubble data and 1048 points Pantheon supernovae data. To study various dark energy models, we use statefinder analysis to address the current cosmic acceleration. We also observe the $Om$ diagnostic describing various phases of the universe. Finally, it is seen that the solution which mimics the power-law fits well with the Pantheon data better than the Hubble data.
} %
\titlerunning{Power-law cosmology in Weyl-type $f(Q,T)$ gravity} 
\authorrunning{G. Gadbail, S. Arora, P.K. Sahoo}
\maketitle
\section{Introduction}

The accelerated expansion becomes a prominent theme in modern cosmology, confirming various observable evidence such as type Ia supernovae observations \cite{Riess/1998,Perlmutter/1999,Spergel/2007}, baryon acoustic oscillations \cite{Eisenstein/2005,Cole/2005}, and large-scale structure \cite{Hawkins/2003}. The incorporation of Riemann geometry into General Relativity (GR), the most successful theory has provided a robust mathematical framework for describing gravitational field properties. However, recent observational data have raised some concerns about the classical GR absolute validity, which may still have some limitations on large or solar system scales. The two significant challenges confronting modern gravitational theories are the dark energy and dark matter problems, which aid in the accelerated expansion of the universe. Modifying the gravitational part of the Einstein's equations is one of the approaches for explaining the acceleration. This approach is named as the modified theory of gravity. So far, several modified theories of gravity beyond GR have been proposed such as the $f(R)$ gravity \cite{Buchdahl/1970,Capozziello/2011,Nojiri/2003}, the $f(R,T)$ gravity \cite{Arora/2020,Harko/2011,Moraes/2017,Yousaf/2018}, the $f(R, G)$ gravity \cite{Atazadeh/2014,Laurentis/2015}, the $f(T,B)$ gravity \cite{Bahamonde/2018,Capo/2020} etc.\\
Within the geometrical framework, the curvature is not the only geometrical object, torsion and non-metricity are the other two fundamental objects related to the connection of a metric space. There are three equivalent representations of GR. The curvature representation in which the torsion and the non-metricity are zero. The second is the teleparallel representation based entirely on the torsion. And there is the symmetric teleparallel representation, in which the non-metricity is associated with gravity.
Weyl suggested a Riemannian extension after GR in an effort to combine gravity and electromagnetism \cite{Weyl/1918}. The non-metricity of spacetime produces the electromagnetic field. Under parallel transport, both the orientation and the length of vectors vary. The dilatational gauge vector, often known as the Weyl vector, is the new vector part of the connection established. There is a scale transformation that transforms this vector to zero when it is given by the gradient of a function. As a result, the lengths of parallel transported vectors through closed pathways return undisturbed, resulting in an integrable Weyl geometry \cite{Scholz/2017,Wheeler/2018}. Higher symmetry techniques to gravity also use Weyl geometry. Moreover, Weizenb$\ddot{o}$ck constructed a geometry with torsion and zero Riemann curvature, another essential mathematical breakthrough with applications \cite{Weitzenbock/1923}. The primary idea behind the teleparallel formulation of gravity is to use tetrad vectors to replace the metric $g_{\mu \nu}$ of the spacetime that describes the gravitational field. It is named as the teleparallel equivalent to GR or $f(T)$ gravity, where $T$ is the torsion \cite{Hayashi/1979}. In recent years, the $f(T)$ theory has yielded some captivating cosmological behaviors, which have been studied in the literature \cite{Benetti/2021,Capozziello/2011a,Mandal/2020,Myrzakulov/2011}. The other geometrically equivalent to GR known as the symmetric teleparallel gravity, was also introduced and further developed into $f(Q)$ gravity \cite{Jimenez/2018}, where the non-metricity $Q$ of a Weyl geometry represents the basic geometrical variable describing the variation of the length of a vector in the parallel transport. Lazkoz et al. \cite{Lazkoz/2019} analysed different forms of $f(Q)$ gravity to study an accelerated expansion of the universe with recent observations. The behavior of cosmological solutions and growth index of matter perturbations in $f(Q)$ gravity has also been investigated in \cite{Khyllep/2021}. Mandal et al. \cite{Mandal/2020a} also studied cosmography in $f(Q)$ gravity. Another recent extension of $f(Q)$ gravity known as $f(Q, T)$ gravity \cite{Xu/2019,2020} includes a non-minimal coupling in the gravitational action, in which the Lagrangian is replaced by an arbitrary function $f$ of the non-metricity $Q$ and the trace of the energy-momentum tensor $T$. Many studies have demonstrated that $f(Q, T)$ gravity is viable option for explaining current cosmic acceleration and can provide a consistent solution to the dark energy problem. Arora et al. \cite{Arora/2021} analysed the feasibility of $f(Q, T)$ gravity by constraining an effective equation of state explaining the dark sector of the universe. $f(Q, T)$ gravity also contributes significantly to gravitational baryogenesis \cite{Bhattacharjee/2020}. The energy conditions in $f(Q, T)$ gravity was studied in \cite{Arora/2021b}
In the context of proper Weyl geometry, Yixin et al. \cite{Yixin/2020b} explored $f(Q, T)$ gravity and adopted the explicit equation for non-metricity $Q$ that follows the non-conservation of the metric tensor divergence. Furthermore, the field equations in the theory were derived using the vanishing scalar curvature condition, which was then applied to the gravitational action via the Lagrange multiplier. The Weyl type $f(Q,T)$ gravity has been found to be an alternate and effective way of describing accelerated and decelerated phases of the universe. Studying variety of functional forms and model parameters, Weyl $f(Q,T)$ could provide a strong alternative to the $\Lambda$CDM, especially giving the late-time de sitter phase generated by Weyl geometry. Yang et al. \cite{Yang/2021} used the Weyl $f(Q, T)$ theory to derive the geodesic and the Raychaudhuri equations .
The analysis here aims to investigate if the Weyl type $f(Q, T)$ gravity can be used to study the accelerated phases of the universe without introducing dark energy. We assumed the case of dust matter i.e. $p=0$ and the linear functional form $f(Q, T)= \alpha Q+ \frac{\beta}{6k^{2}} T$, where $\alpha$ and $\beta$ are model parameters. Also, high precision cosmological data obtained observationally, such as Hubble data and Pantheon samples, have been used to constrain the model parameters. We studied the evolution of the universe using the two statefinder diagnostics : the statefinder diagnostics and the Om diagnostics.\\
The following are the portions of the present article: We presented a broad review of the Weyl type $f(Q,T)$ gravity theory and the gravitational action with its field equations in section \ref{sec2}. In section \ref{sec3}, we used 57 points of the Hubble data points and 1048 Pantheon data points to constrain the model parameters and compared our model with $\Lambda$CDM in error bar plots. In section \ref{sec4}, we observed the behavior of energy density and the statefinder diagnostics. We also presented the geometrical $Om(z)$ diagnostic in section \ref{sec5} to illustrate dark energy models. Section \ref{sec6} includes the summary of our results obtained.

\section{Field equations of the Weyl type $f(Q,T)$ Gravity}\label{sec2}

The action in Weyl type $f(Q,T)$ gravity is given as \cite{Yixin/2020b}

\begin{equation}
\label{1}
S=\int d^4x\,\sqrt{-g}\left[\kappa^2\,f(Q,T)-\frac{1}{4}W_{\,\nu \mu}\,W^{\,\nu \mu}-
\frac{1}{2}m^2\,w_{\nu}\, w^\nu +\mathcal{L}_m\right].
\end{equation}
Imposing the Lagrange multiplier $\lambda$ in the gravitational action, we get

\begin{equation}
 \label{2}
 S=\int d^4x\,\sqrt{-g}\left[ \kappa^2\,f(Q,T)-\frac{1}{4}\,W_{\,\nu \mu}\,W^{\,\nu \mu}-\frac{1}{2}m^2\,w_\nu \,w^\nu+ \right.\\
\left. \lambda\,\left(R+6\,\nabla_\alpha w^\alpha-6\,w_\alpha \,w^\alpha\right)+\mathcal{L}_m\right].
\end{equation}
where, $\kappa^2=\frac{1}{16\pi G}$, $m$ represents the mass of the particle associated to the vector field $w_\mu$, $\mathcal{L}_m$ is the matter Lagrangian, $f$ is an arbitrary function of the non-metricity $Q$ and the trace of the matter-energy-momentum tensor $T$. The second term in the action is the standard kinetic term and the third term is a mass term of the vector field. Also $g=det(g_{\nu \mu})$ and the scalar non-metricity $Q$ is given by 

\begin{equation}
\label{3}
Q\equiv- g^{\nu \mu}\left(L^\alpha_{\,\,\beta\mu}L^\beta_{\,\,\mu\alpha}-L^\alpha_{\,\,\beta\alpha}L^\beta_{\,\,\nu \mu}\right),
\end{equation}
where, $L^\lambda_{\,\,\nu \mu}$ is the deformation tensor defined as
\begin{equation}
\label{4}
L^\lambda_{\,\,\nu \mu}=-\frac{1}{2}g^{\lambda\gamma}\left(Q_{\,\nu\gamma\mu}+Q_{\,\mu\gamma\nu}-Q_{\,\gamma\nu\mu}\right).
\end{equation}
In the Riemannian geometry, the covariant derivative of metric tensor is zero, i.e., $\nabla_\alpha g_{\nu\mu}=0$. But in Weyl geometry, the expression is represented as \cite{Haghani/2012}  
\begin{equation}
\label{5}
Q_{\,\alpha\nu\mu}\equiv\widetilde{\nabla}_\alpha \,g_{\nu\mu}=\partial_\alpha g_{\nu\mu}-\widetilde{\Gamma}^\rho_{\,\,\alpha\nu}\,g_{\rho\mu}-\widetilde{\Gamma}^\rho_{\,\,\alpha\mu}\,g_{\rho\nu}=2w_\alpha\,g_{\nu\mu},
\end{equation}
where, $\widetilde{\Gamma}^\lambda_{\,\,\nu\mu}\equiv\Gamma^\lambda_{\,\,\nu\mu}+g_{\nu\mu}\,w^\lambda-\delta^\lambda_\nu\, w_\mu-\delta^\lambda_\mu\, w_\nu$ and $\Gamma^\lambda_{\,\,\nu\mu}$ is the christoffel symbol with respect to the metric $g_{\nu\mu}$.\\\\
Putting Eq. \eqref{5} in Eq. \eqref{3}, we get the relation 
\begin{equation}
\label{6}
Q=-6w^2 .
\end{equation}
We get the generalized Proca equation explaining the field evolution by varying the action with respect to the vector field,
\begin{equation}
\label{7}
\nabla^{\,\mu} W_{\,\nu\mu}-\left(m^2+12\kappa^2f_Q+12\lambda\right)w_\nu=6\nabla_\nu \lambda .
\end{equation}
We observe that the effective dynamical mass of the vector field when compared with the standard Proca equation read as
\begin{equation}
\label{8}
m^2_{\,eff}=m^2+12\kappa^2f_Q+12\lambda .
\end{equation}
Variation of the gravitational action in Eq. \ref{2} with to the metric tensor gives us the following generalized gravitational field equation:
\begin{multline}
\label{9}
\frac{1}{2}\left(T_{\,\nu\mu}+S_{\,\nu\mu}\right)-\kappa^2 f_{T}\left(T_{\,\nu\mu}+\Theta_{\,\nu\mu}\right)=-\frac{\kappa^2}{2}g_{\nu\mu} \\
-6 \kappa^2f_Q w_\nu w_\mu + \lambda\left(R_{\,\nu\mu}-6\,w_\nu \,w_\mu +3\,g_{\nu\mu}\,\nabla_\rho \,w^\rho \right) \\
+3\,g_{\nu\mu}\,w^\rho \,\nabla_\rho \,\lambda -6\,w_{(\nu}\,\nabla_{\mu )}\,\lambda
+g_{\nu\mu}\,\square \lambda-\nabla_\nu\nabla_\mu \lambda.
\end{multline}
Here, we define,
\begin{equation}
\label{10}
f_T\equiv \frac{\partial f(Q,T)}{\partial T}, \hspace{0.3in}
f_Q\equiv\frac{\partial f(Q,T)}{\partial Q},
\end{equation}

\begin{equation}
\label{11}
T_{\,\nu\mu}\equiv-\frac{2}{\sqrt{-g}}\frac{\delta(\sqrt{-g}\,L_m)}{\delta g^{\nu\mu}},
\end{equation} 
respectively. Also we have defined the quantity $\Theta_{\,\nu\mu}$
\begin{equation}
\label{12}
\Theta_{\,\nu\mu}=g^{\alpha\beta}\,\frac{\delta T_{\,\alpha\beta}}{\delta g_{\mu\nu}}=g_{\nu\mu}\,L_m-2\,T_{\,\nu\mu}-2\,g^{\alpha\beta}\frac{\delta^2 \,L_m}{\delta g^{\nu\mu}\,\delta g^{\alpha\beta}}.
\end{equation}
The re-scaled energy momentum tensor $S_{\nu\mu}$ of the Proca field is defined as
\begin{equation}
\label{13}
S_{\,\nu\mu}=-\frac{1}{4}g_{\nu\mu}\,W_{\,\rho\sigma}\,W^{\,\rho\sigma}+W_{\,\nu\rho}\,W^{\,\rho}_{\,\mu} -\frac{1}{2}m^2\,g_{\nu\mu}\,w_\rho \,w^\rho \\
+m^2\, w_\nu\, w_\mu,
\end{equation}
where 
\begin{equation}
\label{14}
W_{\nu\mu}=\nabla_\mu w_\nu-\nabla_\nu  w_\mu.
\end{equation}

We assume that the FLRW metric in the spatially flat, isotropic and homogeneous universe, given by
\begin{equation}
\label{15}
ds^2=-dt^2+a^2(t) \delta_{ij}  dx^i  dx^j ,
\end{equation}
Where, $a(t)$ is the scale factor. Because of spatial symmetry, the vector field is chosen in the form of
\begin{equation}
\label{16}
w_\nu=\left(\psi (t), 0, 0, 0\right) .
\end{equation}
Using the above equation, we get $w^2=w_\nu w^\nu=-\psi^2(t)$, with $Q=-6w^2=6\psi^2(t)$.\\
So, $u^\nu \nabla_\mu=\frac{d}{dt}$ and $H=\frac{\dot{a}}{a}$. The Lagrangian of the perfect fluid is also assumed to be $\mathcal{L}_m=p$.\\
Now we consider the energy momentum tensor for the perfect fluid is given by 
\begin{equation}
\label{17}
T_{\,\nu\mu}=\left(\rho+p\right)\,u_\nu \,u_\mu+p\,g_{\nu\mu},
\end{equation}
where $\rho$ and $p$ are the energy density and the pressure, respectively, $u^\nu$ is the four velocity vector satisfying the condition $u_\nu \,u^\nu=-1$. Hence, we have
\begin{equation}
\label{18}
 T^{\,\nu}_{\,\mu}=diag\left(-\rho, p, p, p\right),
\end{equation}
and
\begin{equation}
\label{19}
\Theta^{\,\nu}_{\,\mu}=\delta^{\,\nu}_{\,\mu} p-2T^{\,\nu}_{\,\mu}=diag\left(2 \rho+p, -p,  -p, -p\right).
\end{equation}
In cosmological case, the constraint of flat space and the generalized Proca equation are obtained as
\begin{equation}
\label{20}
\dot{\psi}=\dot{H}+2 H^2+\psi^2-3 H \psi,
\end{equation}
\begin{equation}
\label{21}
\dot{\lambda}=\left(-\frac{1}{6}m^2-2\kappa^2 f_Q-2\,\lambda\right) \psi=-\frac{1}{6} m^2_{ eff} \psi ,
\end{equation}
\begin{equation}
\label{22}
\partial_i \lambda=0
\end{equation}
From Eq. \eqref{9} the generalized Friedmann equation read as
\begin{equation}
\label{23}
\kappa^2f_T\,\left(\rho+p\right)+\frac{1}{2}\rho=\frac{\kappa^2}{2}f-\left(6\kappa^2f_Q+\frac{1}{4}m^2\right)\psi^2 \\
-3\lambda\left(\psi^2-H^2\right)-3\dot{\lambda}\left(\psi-H\right),
\end{equation}

\begin{equation}
\label{24}
-\frac{1}{2}p=\frac{\kappa^2}{2}f+\frac{m^2\psi^2}{4}+\lambda\left(3\psi^2+3H^2+2\dot{H}\right)\\
+\left(3\psi+2H\right)\dot{\lambda}+\ddot{\lambda}.
\end{equation}
where dot$(\cdot)$ represents the derivative with respect to time and $f_Q$ and $f_T$ represent differentiation with respect to $Q$ and $T$ respectively.\\

Now, we consider the functional form $f(Q,T)=\alpha \,Q+\frac{\beta}{6\,\kappa^2}\,T$, where $\alpha$ and $\beta$ are model parameters. The functional form depends on three free parameters $\alpha$, $\beta$ and $M^2=\frac{m^2}{\kappa^2}$, $M$ is the mass of the Weyl field, indicating the strengths of the Weyl geometry-matter coupling. In this case, we have assumed $M=0.95$. It is worth mentioning that $\beta=0$ corresponds to the $f(Q, T)= \alpha Q$ i.e. a case of the successful theory of General Relativity (GR). Also,  $T=0$, the case of vacuum, the theory reduces to $f(Q)$ gravity, which is equivalent to GR, that passes all Solar System tests, considered in the vacuum.  Furthermore, Yixin et al. \cite{Xu/2019,2020} also depicts that the universe experiences an accelerating expansion ending with a de Sitter type evolution in the considered model. We study the model for bulk viscous fluid in a non relativistic case, i.e. $p=0$. 
Using this in Eq. \eqref{23} and \eqref{24} we get,

\begin{multline}
\label{25}
\left(12\kappa^2 \alpha+m^2+\frac{18}{\beta}\,\kappa^2\alpha+\frac{3}{2\beta}m^2 \right)\psi^2+3\lambda\left(4\psi^2+2H^2+2\dot{H}\right)
+3\dot{\lambda}\left(4\psi+H\right)+\frac{6\lambda}{\beta}\left(3\psi^2+3H^2+2\dot{H}\right)+\\\frac{6\dot{\lambda}}{\beta}\left(3\psi+2H\right)+3\ddot{\lambda}
+\frac{6}{\beta}\ddot{\lambda}=0,
\end{multline}
where $\alpha$ and $\beta$ are constants and $T=3\,p-\rho$.\\
We try to solve the above equation by considering $\lambda=\kappa^2$,

\begin{equation}
\label{26}
\left(12\,\kappa^2 \alpha+m^2+\frac{18}{\beta}\kappa^2\alpha+\frac{3}{2\beta}m^2 \right)\psi^2+\\
3\kappa^2\left(4\psi^2+2H^2+2\dot{H}\right)
+\frac{6\kappa^2}{\beta}\left(3\psi^2+3H^2+2\dot{H}\right)=0.
\end{equation}
Further simplifying the above equation,
\begin{equation}
 \left(\left(12\beta+18\right)\alpha+\left(\beta+\frac{3}{2}\right)M^2\right)\psi^2+\\
 3\beta \left(4\psi^2+2H^2+2\dot{H}\right)
 +6\left(3\psi^2+3H^2+2\dot{H}\right)=0
 \label{27}
\end{equation}

Using the relation $\nabla_\lambda g_{\nu\mu}=-w_\lambda g_{\nu\mu}$ and Eq. \eqref{16}, we obtain $\psi= -6H$ and Eq. \eqref{27} read as
\begin{equation}
\label{28}
A\,H^2+B\,\dot{H}=0
\end{equation}
where,
 $A=36\left( \left(12\beta+18\right)\alpha+\left(\beta+\frac{3}{2}\right)M^2+\left(12\beta+18\right)\right) +\left(6\beta+18\right)$ and $B=\left(6\beta+12\,\right)$\\
 The obtained solution of differential equation given in \eqref{28} is,
 
\begin{equation}
\label{29}
 H(t)=\frac{1}{k_1\,t-c_1}
\end{equation}
where,
\begin{eqnarray*}
 k_{1}=\frac{A}{B}=\left[\frac{36\left(2\beta+3\right)\left(\alpha+1\right)}{\left(\beta+2\right)}+\frac{\left(6\beta+9 \right)}{\left(\beta+2\,\right)}M^2+\frac{\left(\beta+3\right)}{\left(\beta+2\right)}\right]
\end{eqnarray*}
Solving eq. \eqref{29}, the expression of the scale factor is obtained as 
\begin{equation}
\label{30}
a(t)=c_2\left(k_{1} t-c_{1} \right)^\frac{1}{k_1}
\end{equation}
We shall write all cosmological parameter in term of redshift using the relation (taking $a(t_{0})=1$)
\begin{equation}
\label{31}
a(t)=\frac{1}{1+z}
\end{equation}
The Hubble parameter and deceleration parameter in terms of redshift are,
\begin{equation}
\label{32}
H(z)= H_{0}(1+z)^{k_{1}}
\end{equation}
\begin{equation}
\label{33}
q(z)=\left(k_1-1\right)
\end{equation}
Here, we have obtained the power-law as the solutions of the field equations. Power-law cosmology is an intriguing solution for dealing  with some unusual challenges like flatness, horizon problem, etc. The power-law is well-motivated in literature. Kumar \cite{Kumar/2012}  used power-law with Hz and SNe Ia data to analyse cosmological parameters. Rani et al \cite{Rani/2015} also examined the power-law cosmology with statefinder analysis.

\section{Data Interpretation}\label{sec3}

\subsection{Hubble data}

\begin{figure}[H]
\centering
\includegraphics[scale=0.5]{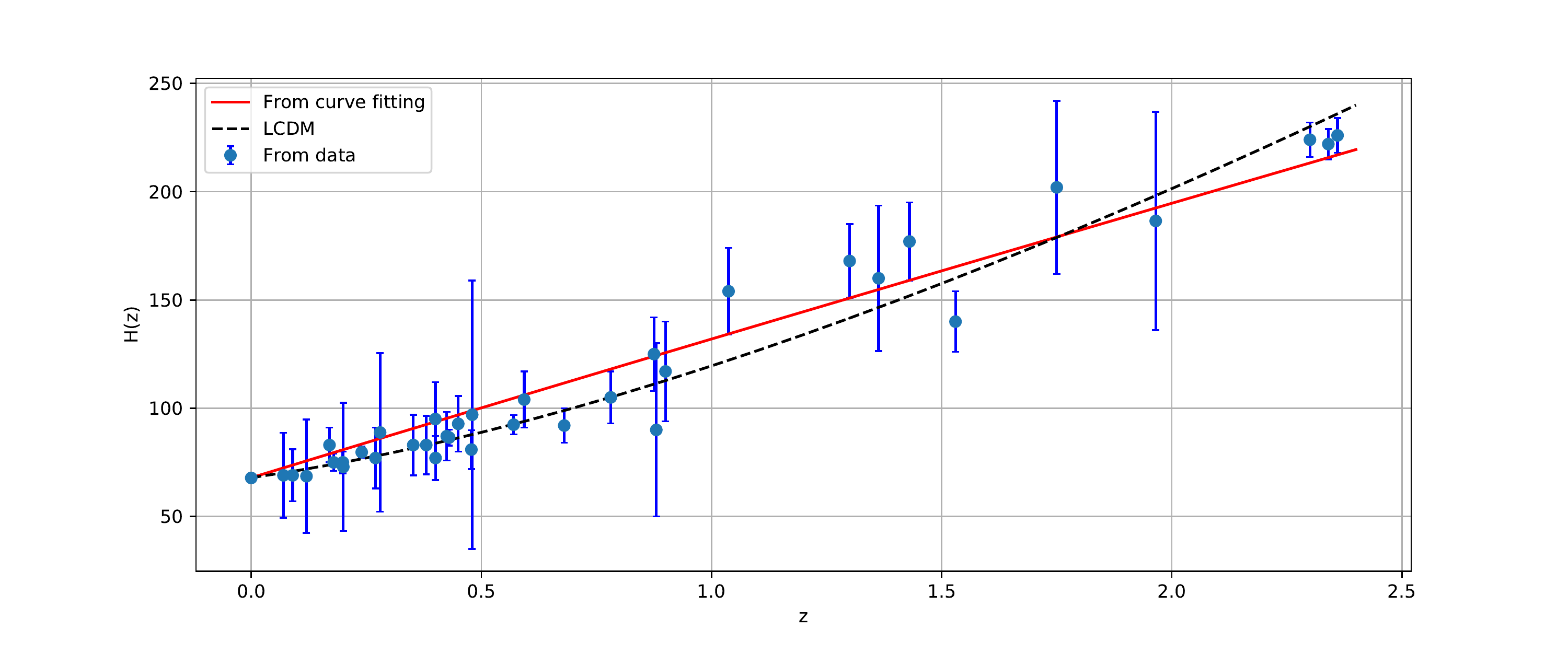}
\caption{The evolution of Hubble parameter with respect to redshift z. The blue dots represents error bars of 57 points, the red line is the curve obtained for our model while black dashed line corresponds to $\Lambda$CDM model. }
\label{fig-1}
\end{figure}

\begin{figure}[H]
\centering
\includegraphics[scale=0.9]{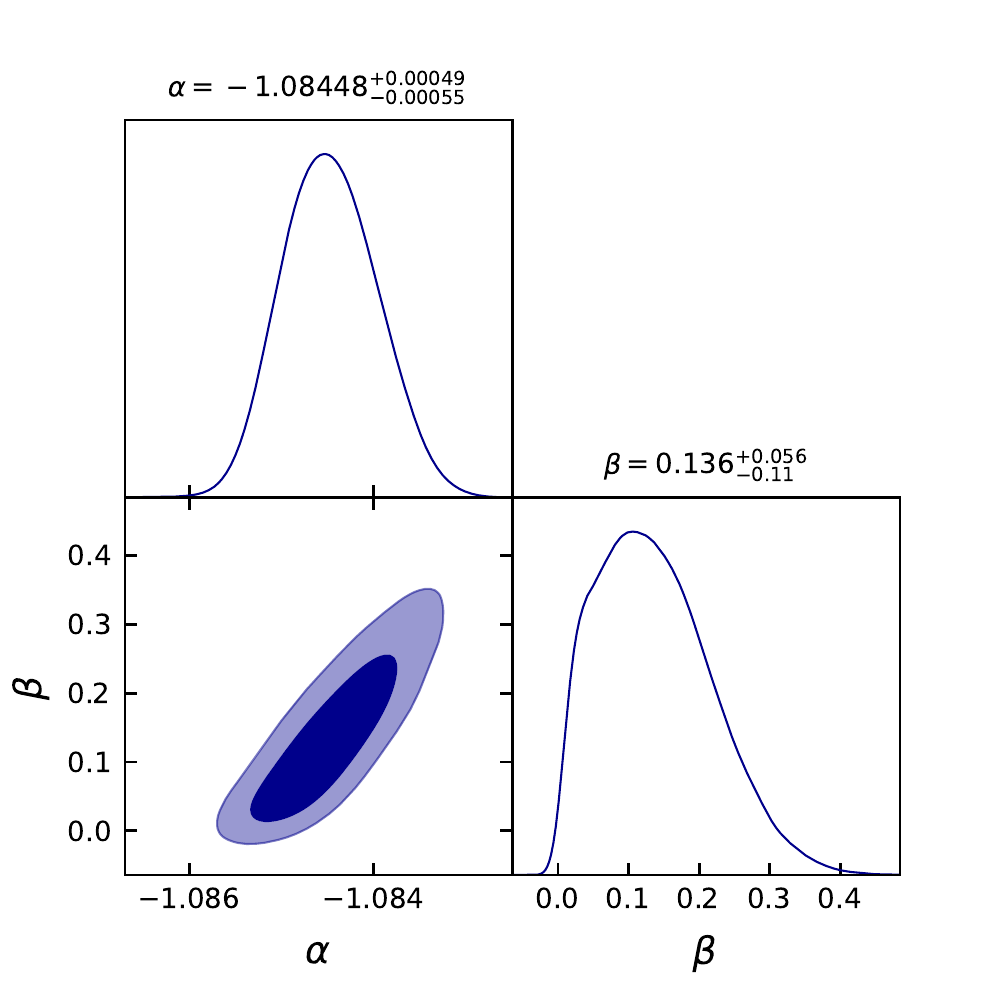}
\caption{The contour plot for model parameters $\alpha$ and $\beta$ with $1-\sigma$ and $2-\sigma$ confidence regions. It also mentions the best fit values of $\alpha$ and $\beta$ using 57 Hubble data points.}
\label{fig-2}
\end{figure}

Numerous observations such as the cosmic microwave background (CMB) form the Wilkinson Microwave Anisotropy Probe team \cite{Hinshaw/2013,Komatsu/2011,Spergel/2007} and Planck team \cite{Planck/2015,Planck/2018}, baryonic acoustic oscillations (BAO) \cite{Eisenstein/2005,Percival/2010}, Type Ia supernovae (SNeIa) \cite{Perlmutter/1999,Riess/1998} have been used to constrain cosmological parameters. Many of these models rely on values that require Hubble parameter to be integrated along the line of sight (the luminosity distance in SNe observations) to explore overall expansion through time. The Hubble parameter $H$ is intimately tied to the expansion history of the universe and is defined as $H=\frac{\dot{a}}{a}$, where $a$ signifies the cosmic scale factor and $\dot{a}$ as the rate of change about cosmic time. The expansion rate $H(z)$ is obtained as 
\begin{equation}
H(z)= -\frac{1}{1+z} \frac{dz}{dt},
\end{equation}
where $z$ is the redshift. 
Two procedures are commonly employed to estimate the value of the $H(z)$ at a certain redshift. One way is to extract $H(z)$ from line-of-sight BAO data, while another uses differential age methods. We used the revised set of 57 data points, which comprises 31 points from the differential age (DA) approach and the left 26 points measured using BAO and other redshift range $0.07 < z< 2.42$. In addition for our investigation, we use $H_{0}= 67.8 km s^{-1} Mpc^{-1}$. The chi-square function is defined to find the mean values of the model parameters $\alpha$ and $\beta$.
\begin{equation}
\chi^{2}_{Hubb} (\alpha, \beta)= \sum_{i=1}^{57} \dfrac{\left[ H^{th}_{i}(\alpha,\beta,z_{i})-H^{obs}_{i}(z_{i})\right]^{2} }{\sigma^{2}(z_{i})}
\end{equation}
where $H^{obs}_{i}$ denotes the observed value, $H^{th}_{i}$ indicates the Hubble's theoretical value while the standard error in the observed value is denoted by $\sigma(z_{i})$.
We used error bars to represent 57 points of $H(z)$ and compared our model with the well-accepted $\Lambda$CDM model in fig. \ref{fig-1}. We considered $H_{0}=67.8 km s^{-1} Mpc^{-1}$, $\Omega_{\Lambda_{0}}=0.7$ and $\Omega_{m_{0}}=0.3$. The best fit values of $\alpha$ and $\beta$ are obtained through data as shown in triangle plot \ref{fig-2} with $1-\sigma$ and $2-\sigma$ confidence intervals. The bounds from our analysis are $\alpha= -1.08448^{+0.00049}_{-0.00055}$ and $\beta=0.136^{+0.056}_{-0.110}$.


\subsection{Pantheon data}

\begin{figure}[H]
\centering
\includegraphics[scale=0.5]{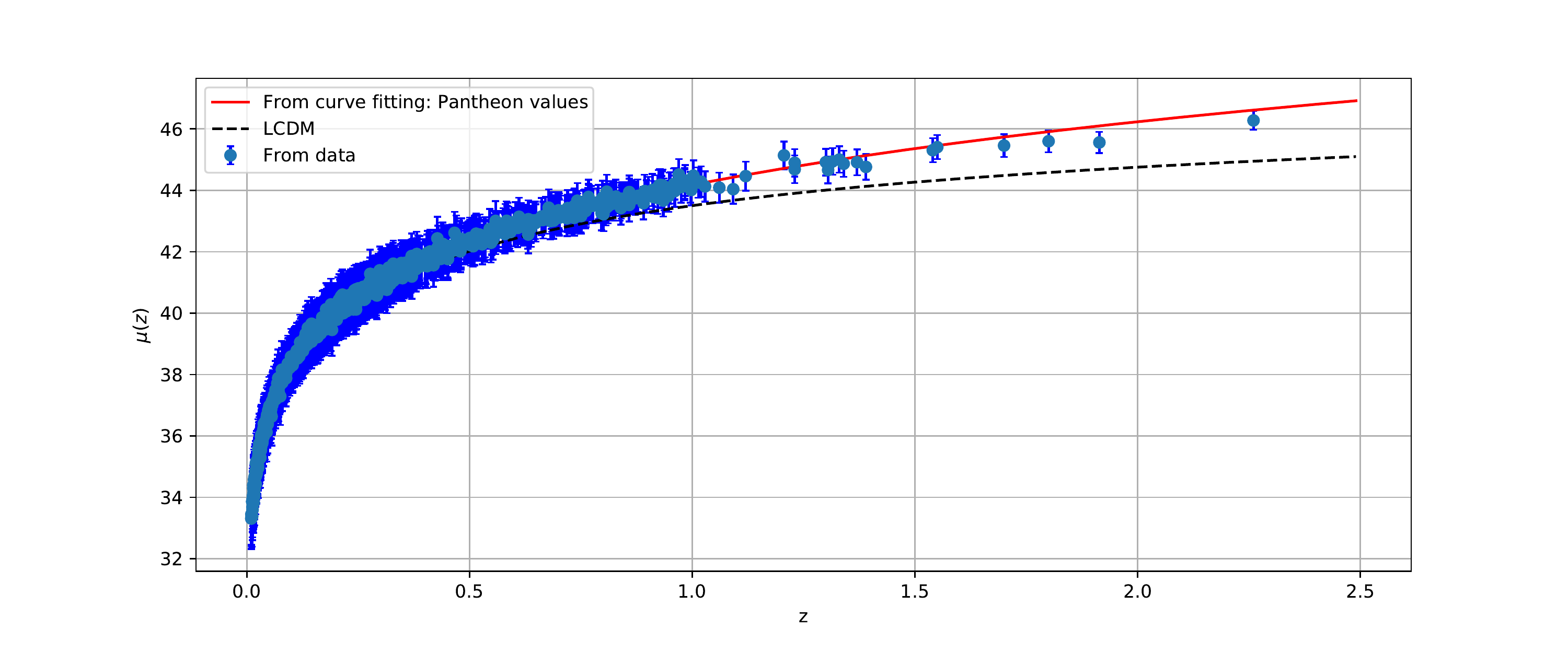}
\caption{The plot of $\mu(z)$ with respect to redshift $z$. The blue dots represents error bars of 1048 pantheon points, the red line is the curve obtained for our model while black dashed line corresponds to $\Lambda$CDM model.}
\label{fig-3}
\end{figure}

\begin{figure}[H]
\centering
\includegraphics[scale=1.0]{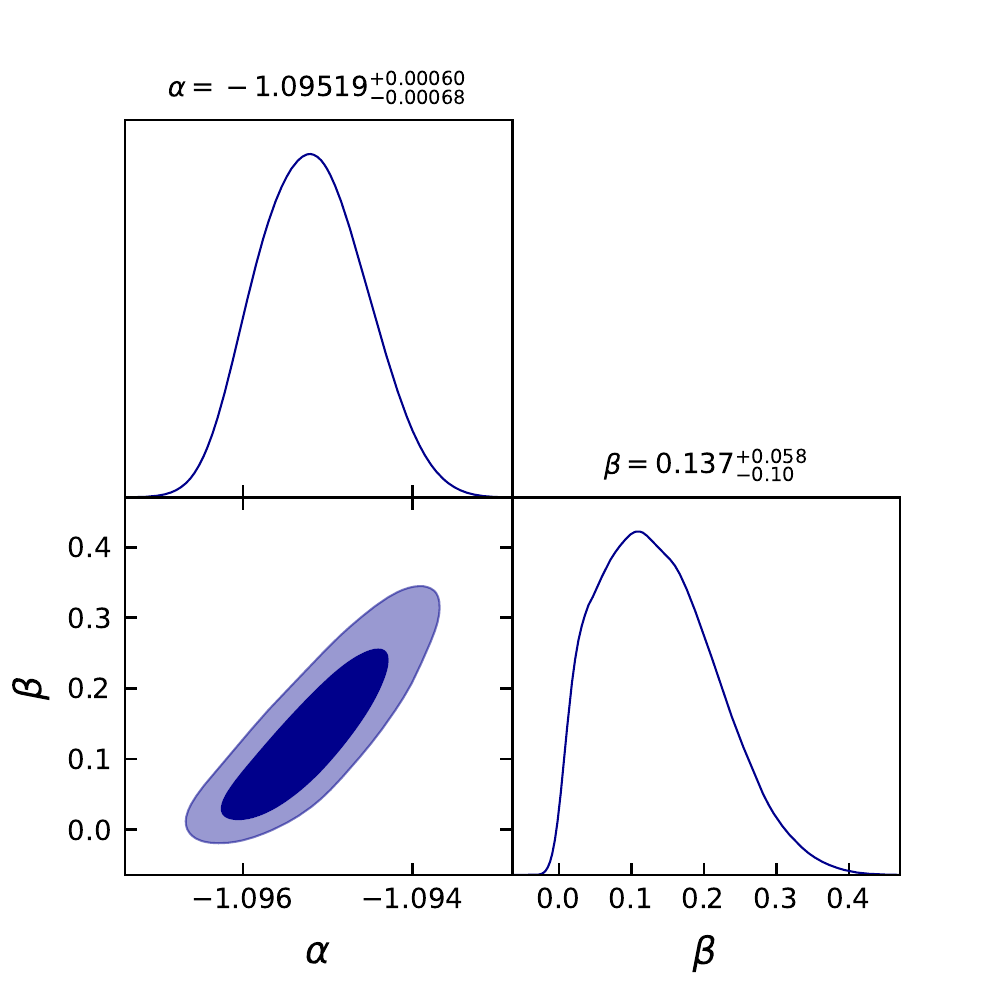}
\caption{The contour plot for model parameters $\alpha$ and $\beta$ with $1-\sigma$ and $2-\sigma$ confidence regions. It also mentions the best fit values of $\alpha$ and $\beta$ using 1048 pantheon samples.}
\label{fig-4}
\end{figure}

We use the most recent compilation of Supernovae pantheon samples to constrain the model parameters $\alpha$ and $\beta$. The Pantheon sample consists of 1048 SNe Ia in the range of $0.01<z<2.26$ \cite{Camlibel/2020,Scolnic/2018}. The likelihood function is determined using the MCMC approach and emcee Python's library to calculate the posterior distributions of the model parameters. The pantheon data are shown in $(m, z)$ pairs, with $m$ typically to be measured. 
The theoretical distance modulus is defined as 
\begin{equation}
\mu^{th}= 5 log_{10} \left( \frac{D_{L} H_{0}^{-1}}{Mpc}\right)  +25,
\end{equation}
where we define
\begin{equation}
D_{L}= (1+z) c \int_{0}^{z} \frac{d\bar{z}}{H(\bar{z}) }.
\end{equation}
Here $H_{0}$ is the Hubble constant. The chi-square function according to our considered model is given as 
\begin{equation}
\chi^{2}_{Pan}(\alpha,\beta)= \sum_{i=1}^{1048} \dfrac{\left[ \mu^{th}_{i}(\alpha,\beta,z_{i})-\mu^{obs}_{i}(z_{i})\right]^{2} }{\sigma^{2}(z_{i})}
\end{equation}
where $\sigma^{2}(z_{i})$ is the standard error, $\mu^{th}_{i}= m-M$ is the theoretical value with $m$ and $M$ are the apparent and absolute magnitudes respectively, $\mu^{obs}_{i}$ is the observed values from data points.
We used error bars to represent 1048 points of pantheon samples and compared our model with the well-accepted $\Lambda$CDM model in fig. \ref{fig-3}. We considered $H_{0}=67.8 km s^{-1} Mpc^{-1}$, $\Omega_{\Lambda_{0}}=0.7$ and $\Omega_{m_{0}}=0.3$. The best fit values of $\alpha$ and $\beta$ are obtained through pantheon samples as shown in triangle plot \ref{fig-4} with $1-\sigma$ and $2-\sigma$ confidence intervals. The bounds from our analysis are $\alpha=-1.09519^{+0.00060}_{-0.00068}$ and $\beta=0.137^{+0.058}_{-0.100}$.

\section{Cosmological parameters}\label{sec4}

\subsection{Density parameter}

By solving eqs. \eqref{23} and \eqref{24}, we can obtain an expression for the density parameter $\rho$. The behavior of density parameter is shown below in fig. \ref{fig-5} and \ref{fig-6} for the obtained $\alpha$ and $\beta$ from Hubble and Pantheon datasets respectively. It can be observed that the density parameter for both the datasets is showing a positive behavior with redshift $z$.

\begin{figure}[H]
\centering
\includegraphics[scale=0.45]{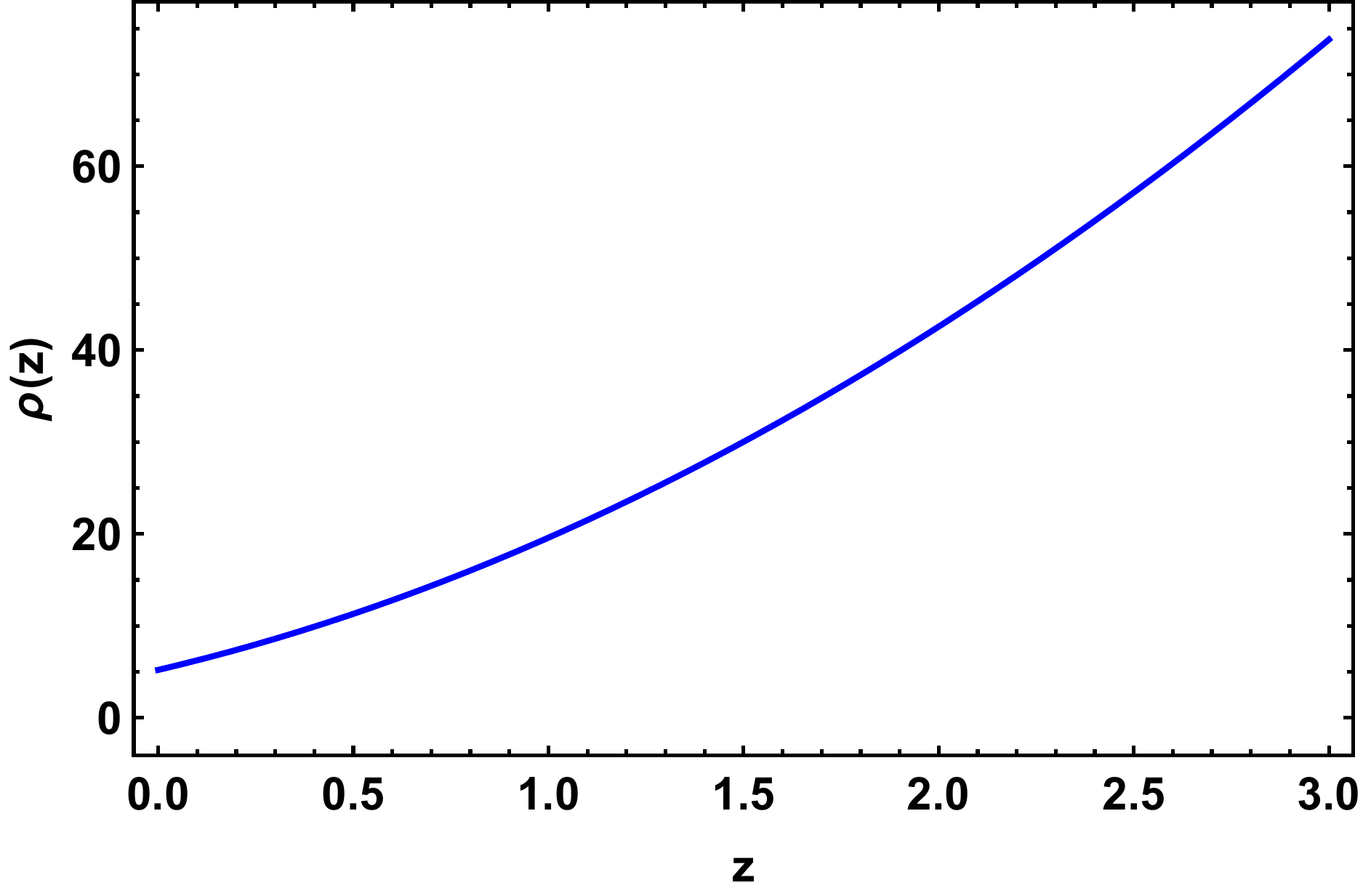}
\caption{The behavior of density parameter for $\alpha= -1.08448$ and $\beta=0.136 $ obtained from Hubble data.}
\label{fig-5}
\end{figure}

\begin{figure}[H]
\centering
\includegraphics[scale=0.45]{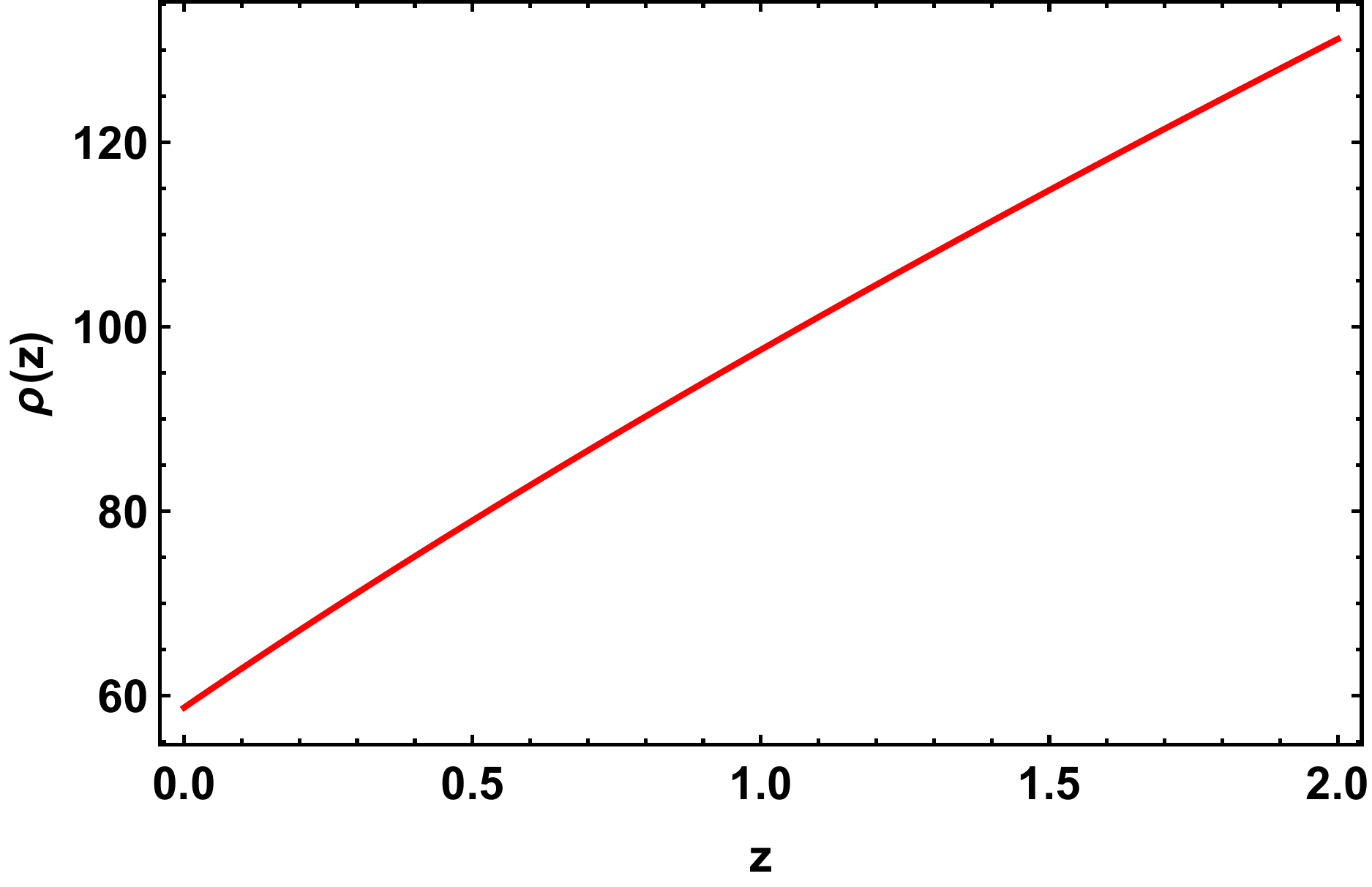}
\caption{The behavior of density parameter for $\alpha=-1.09519$ and $\beta= 0.137$ obtained from Pantheon data.}
\label{fig-6}
\end{figure}

\subsection{Statefinder diagnostics}

\begin{figure}[H]
\centering
\includegraphics[scale=0.45]{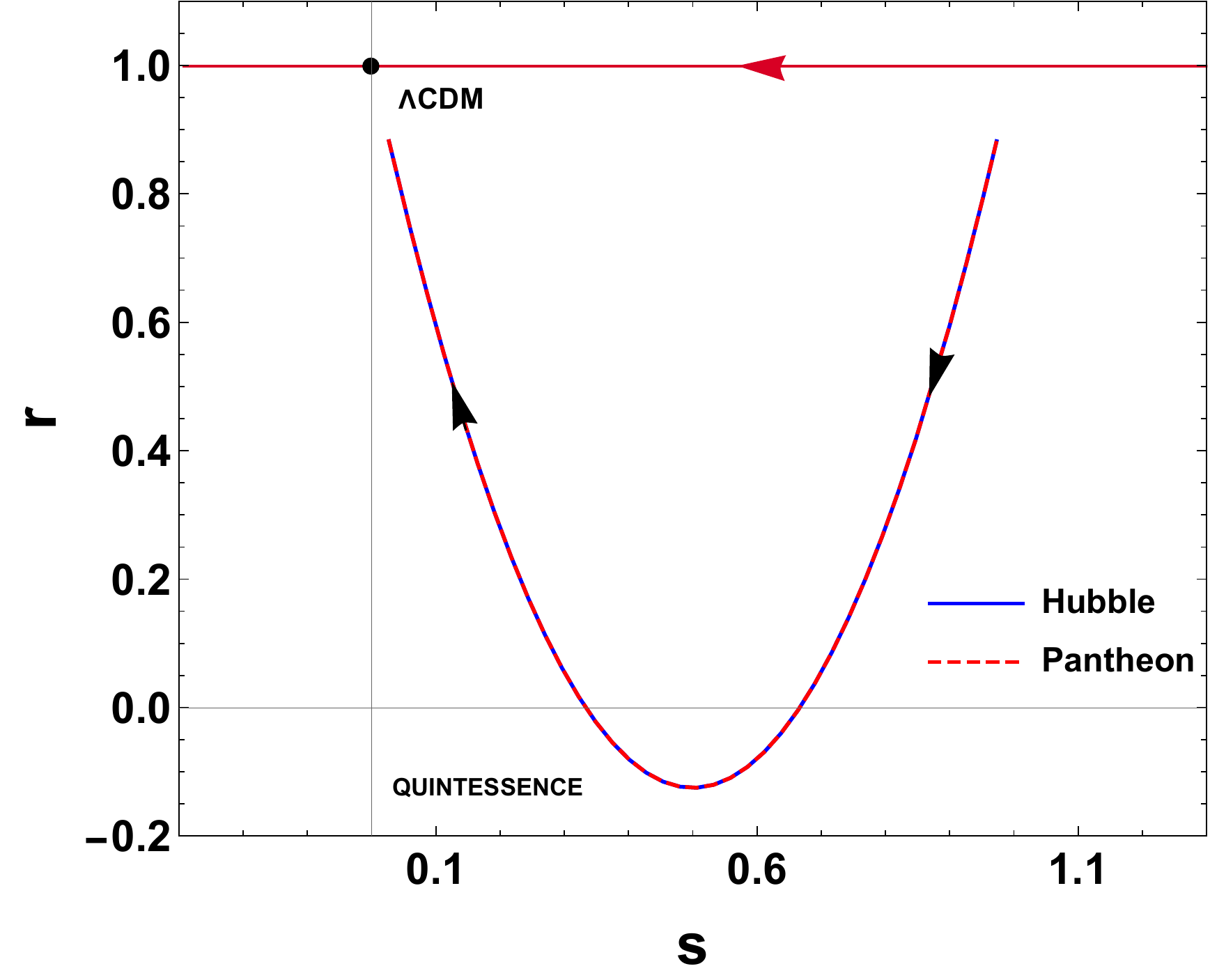}
\caption{The figure shows the behavior of $s-r$ plane with $\beta=0.136$ and $\beta=0.137$ and varied $\alpha$.}
\label{fig-7}
\end{figure}

\begin{figure}[H]
\centering
\includegraphics[scale=0.45]{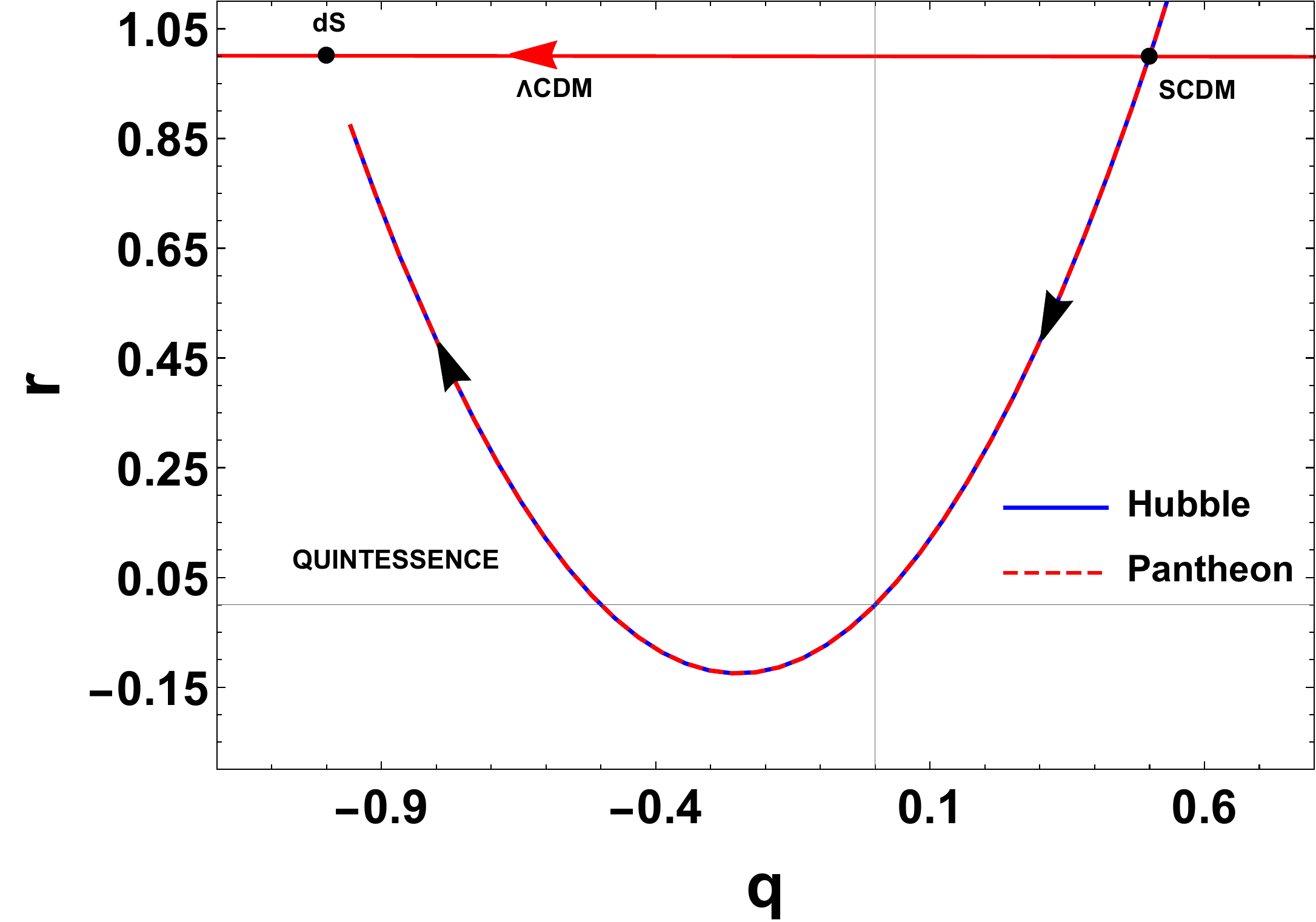}
\caption{The figure shows the graph of $q-r$ plane with $\beta=0.136$ and $\beta=0.137$ and varied $\alpha$.}
\label{fig-8}
\end{figure}

Numerous DE models can be used to describe cosmic acceleration. Another reliable diagnostic exists to distinguish between many cosmological models involving dark energy. Sahni et al. \cite{Sahni/2003,Alam/2003} proposed a new dark energy diagnostic known as  statefinder diagnostics, dependent on the second and third derivatives of the scale factor. It is defined with the help of well known geometrical parameters namely the Hubble parameter $H=\frac{\dot{a}}{a}$ and the deceleration parameter $q=-\frac{\ddot{a}}{aH^2}$. The statefinder parameter pair ${r-s}$ is defined as

\begin{equation}
\label{36}
r=\frac{\dddot{a}}{a \,H^3}
\end{equation}
\begin{equation}
\label{37}
s=\frac{r-1}{3\,(q-\frac{1}{2})}
\end{equation}
The plot of ${s-r}$ is shown in fig. \ref{fig-7}. The statefinder parameter ${s-r}$ can be an admirable diagnostic  for describing significant dark energy model characteristics. According to the trajectories in ${s-r}$ plane, the point $(0,1)$ corresponds to the $\Lambda$CDM model, Chaplygin gas lie to the left of the $\Lambda$CDM model whereas quintessence lie  to the right of the $\Lambda$CDM. The evolution of ${q-r}$ is shown in Fig. \ref{fig-8}. It is observed the point $(q, r)=(0.5, 1)$ correspond to $SCDM$ (i.e. matter dominated universe), with the de-sitter (dS) expansion pointing to $(q, r) =(-1, 1)$ in the future. As a result, the statefinder diagnostics can successfully distinguish between various dark energy models.

It is worth noting that in the obtained model, the $\Lambda$CDM statefinder pair $(0,1)$ and correspondingly the dS point $(-1,1)$ acts as an attractor. The constraints on statefinder from Hubble data and Pantheon data are obtained as $r= -0.039^{+0.009}_{-0.009}$, $s= 0.637^{+0.0075}_{-0.0075}$ and $r= -0.105^{+0.008}_{-0.011}$, $s= 0.434^{+0.023}_{-0.012}$ respectively \cite{Kumar/2012,Rani/2015}. It is observed that the model fits well with Pantheon datasets rather than the Hubble data. 

\section{Om Diagnostics} \label{sec5}

\begin{figure}[H]
\centering
\includegraphics[scale=0.5]{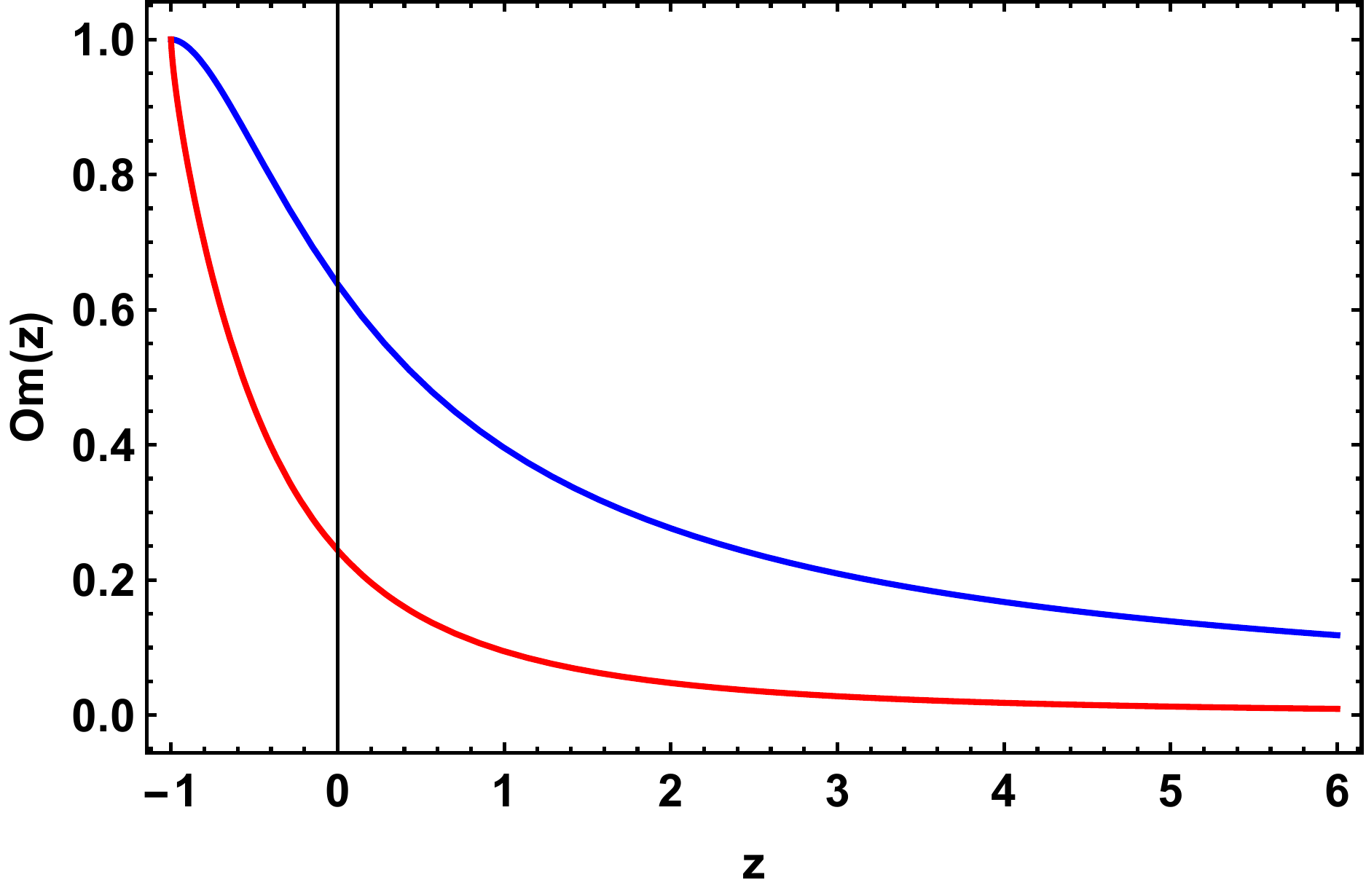}
\caption{This figure shows the behavior of $Om$ vs redshift $z$ with $\alpha=-1.08448$, $\beta=0.136$ and $\alpha=-1.09519$, $\beta=0.137$ constrained by Hubble and Pantheon datasets respectively.} \label{fig-9}
\end{figure}

The $Om$ diagnostic can be studied as a simplest diagnostic than the statefinder diagnostic \cite{Sahni/2008,Shahalam/2015} because it uses only the first-order time derivative of scale factor i.e involving the Hubble parameter. It is used to clarify various dark energy (DE) models by differentiating $\Lambda$ CDM model. For spatially flat universe, the $Om(z)$ diagnostic is defined as
\begin{equation}
Om\left(z\right)=\frac{\left(\frac{H\left(z\right)}{H_0}\right)^2-1}{\left(1+z\right)^3-1}
\end{equation} 
where, $H_0$ is the Hubble constant.
According to the behavior of $Om(z)$, different dark energy models can be described. Phantom type i.e. $\omega < -1$ corresponds to the positive slope of $Om(z)$, quintessence type $\omega > -1$ corresponding to negative slope of $Om(z)$. The constant behavior of $Om(z)$ depicts the $\Lambda$CDM model.
In Fig. \ref{fig-9}, the $Om(z)$ has a negative slope, showing quintessence-like behavior indicating the accelerated expansion. As a result, the model may not resolve the Hubble tension at present. The study in references \cite{Vagnozzi/2020,Valentino/2016} reveals that a phantom like component with effective equation of state $\omega = -1.29$ can solve the current tension between the Planck data set and other prior in an extended $\Lambda$CDM scenario. It is also worth noting from \cite{Valentino/2021} that the lower tension is attributable to a change in the value of $H_{0}$ and an increase in its uncertainty owing to degeneracy with more physics, further confounding the picture and indicating the need for more probes. While no single idea stands out as very plausible or superior to all other, solutions including early or dynamical dark energy, interacting cosmologies and modified gravity are the best alternatives until a better one emerges.

\section{Conclusion}\label{sec6}

In this study, we considered an extension of the third equivalent representation of GR (the symmetric teleparallel formulation) called $f(Q, T)$ gravity, where the non-metricity $Q$ is non-minimally coupled to the trace $T$ of energy-momentum tensor. We examined the Weyl type $f(Q, T)$ gravity, in which the product of the metric and the Weyl vector determines the covariant divergence of the metric tensor. As a result, the Weyl vector and metric tensor is responsible for the geometrical features of the theory. We have considered the case of dust and obtained the solutions  of the field equations. The Hubble parameter is found to be similar to the power-law form in redshift $z$. 
We used the most recent 57 Hubble data sets and 1048 Pantheon supernovae datasets to constrain the model parameters $\alpha$ and $\beta$. The model is also compared to $\Lambda$CDM model shown in the error bar plots. According to the constraints values of $\alpha$ and $\beta$, the deceleration parameter $q$ is seen to be negative. The nature of cosmic evolution in the Weyl $f(Q, T)$ gravity is greatly reliant on the values of the functional form of $f$ and the model parameters involved. As a result, we used the statefinder diagnostics $s-r$ and $q-r$ and the $Om$ diagnostic analysis for the model to study the nature of dark energy models. The constrained values of $r$ and $s$ are obtained as $r= -0.039^{+0.009}_{-0.009}$, $s= 0.637^{+0.0075}_{-0.0075}$ and $r= -0.105^{+0.008}_{-0.011}$, $s= 0.434^{+0.023}_{-0.012}$ for Hubble and Pantheon data respectively. It is observed that the model fits well with Pantheon $SNe Ia$ data better that the $Hz$ data. The obtained model is proven to be helpful in describing the acceleration of present universe in the context of current observations of $Hz$ and $SNe Ia$. However, it fails to provide redshift transition from deceleration to acceleration due to the constant value of the deceleration parameter. Hence, there are many other possibilities to check the viability of Weyl $f(Q, T)$ theory, such as considering of the scalar field to study inflation, a theoretical study in the presence of coupling between geometry and matter, etc.

\section*{Acknowledgments}

GG RS acknowledges University Grants Commission (UGC), New Delhi, India for awarding Junior Research Fellowship (UGC-Ref. No.: 201610122060). SA acknowledges CSIR, Govt. of India, New Delhi, for awarding Junior Research Fellowship. PKS acknowledges CSIR, New Delhi, India for financial support to carry out the Research project [No.03(1454)/19/EMR-II Dt.02/08/2019]. We are very much grateful to the honorable referee and the editor for the illuminating suggestions that have significantly improved our work in terms of research quality and presentation.

\section{Data Availability Statement}
There are no new data associated with this article.

\section{Appendix}
Here, Table-1 contains the $57$ points of Hubble parameter values $H(z)$ with errors $\sigma _{H}$ from differential age ($31$ points), and BAO and other ($26$ points) approaches, along with references.

\begin{center}
\begin{tabular}{|c|c|c|c|c|c|c|c|}\hline
\multicolumn{8}{|c|}{Table-1: $H(z)$ datasets consisting of 57 data points} \\ \hline
\multicolumn{8}{|c|}{DA method (31 points)}  \\ \hline
$z$ & $H(z)$ & $\sigma _{H}$ & Ref. & $z$ & $H(z)$ & $\sigma _{H}$ & Ref. \\ \hline
$0.070$ & $69$ & $19.6$ & \cite{h1} & $0.4783$ & $80$ & $99$ & \cite{h5} \\ \hline
$0.90$ & $69$ & $12$ & \cite{h2} & $0.480$ & $97$ & $62$ & \cite{h1} \\ \hline
$0.120$ & $68.6$ & $26.2$ & \cite{h1} & $0.593$ & $104$ & $13$ & \cite{h3} \\ \hline
$0.170$ & $83$ & $8$ & \cite{h2} & $0.6797$ & $92$ & $8$ & \cite{h3} \\ \hline
$0.1791$ & $75$ & $4$ & \cite{h3} & $0.7812$ & $105$ & $12$ & \cite{h3} \\ \hline
$0.1993$ & $75$ & $5$ & \cite{h3} & $0.8754$ & $125$ & $17$ & \cite{h3} \\ \hline
$0.200$ & $72.9$ & $29.6$ & \cite{h4} & $0.880$ & $90$ & $40$ & \cite{h1} \\ \hline
$0.270$ & $77$ & $14$ & \cite{h2} & $0.900$ & $117$ & $23$ & \cite{h2} \\ \hline 
$0.280$ & $88.8$ & $36.6$ & \cite{h4} & $1.037$ & $154$ & $20$ & \cite{h3} \\ \hline 
$0.3519$ & $83$ & $14$ & \cite{h3} & $1.300$ & $168$ & $17$ & \cite{h2} \\ \hline 
$0.3802$ & $83$ & $13.5$ & \cite{h5} & $1.363$ & $160$ & $33.6$ & \cite{h7} \\ \hline 
$0.400$ & $95$ & $17$ & \cite{h2} & $1.430$ & $177$ & $18$ & \cite{h2} \\ \hline 
$0.4004$ & $77$ & $10.2$ & \cite{h5} & $1.530$ & $140$ & $14$ & \cite{h2} \\ \hline
$0.4247$ & $87.1$ & $11.2$ & \cite{h5} & $1.750$ & $202$ & $40$ & \cite{h2} \\ \hline
$0.4497$ & $92.8$ & $12.9$ & \cite{h5} & $1.965$ & $186.5$ & $50.4$ & \cite{h7}  \\ \hline
$0.470$ & $89$ & $34$ & \cite{h6} &  &  &  &   \\ \hline
\multicolumn{8}{|c|}{From BAO \& other method (26 points)} \\ \hline
$z$ & $H(z)$ & $\sigma _{H}$ & Ref. & $z$ & $H(z)$ & $\sigma _{H}$ & Ref. \\ \hline
$0.24$ & $79.69$ & $2.99$ & \cite{h8} & $0.52$ & $94.35$ & $2.64$ & \cite{h10} \\ \hline
$0.30$& $81.7$ & $6.22$ & \cite{h9} & $0.56$ & $93.34$ & $2.3$ & \cite{h10} \\ \hline
$0.31$ & $78.18$ & $4.74$ & \cite{h10} & $0.57$ & $87.6$ & $7.8$ & \cite{h14} \\ \hline
$0.34$ & $83.8$ & $3.66$ & \cite{h8} & $0.57$ & $96.8$ & $3.4$ & \cite{h15} \\ \hline
$0.35$ & $82.7$ & $9.1$ & \cite{h11} & $0.59$ & $98.48$ & $3.18$ & \cite{h10} \\ \hline
$0.36$ & $79.94$ & $3.38$ & \cite{h10} & $0.60$ & $87.9$ & $6.1$ & \cite{h13} \\ \hline
$0.38$ & $81.5$ & $1.9$ & \cite{h12} & $0.61$ & $97.3$ & $2.1$ & \cite{h12} \\ \hline
$ 0.40$ & $82.04$ & $2.03$ & \cite{h10} & $0.64$ & $98.82$ & $2.98$ & \cite{h10}  \\ \hline
$0.43$ & $86.45$ & $3.97$ & \cite{h8} & $0.73$ & $97.3$ & $7.0$ & \cite{h13} \\ \hline
$0.44$ & $82.6$ & $7.8$ & \cite{h13} & $2.30$ & $224$ & $8.6$ & \cite{h16} \\ \hline
$0.44$ & $84.81$ & $1.83$ & \cite{h10} & $2.33$ & $224$ & $8$ & \cite{h17} \\ \hline
$0.48$ & $87.79$ & $2.03$ & \cite{h10} & $2.34$ & $222$ & $8.5$ & \cite{h18} \\ \hline
$0.51$ & $90.4$ & $1.9$ & \cite{h12} & $2.36$ & $226$ & $9.3$ & \cite{h19} \\ \hline
\end{tabular}
\end{center}






\end{document}